\RequirePackage{fix-cm}
\documentclass[smallextended]{svjour3wide} % onecolumn (wide varsion)
\smartqed  % flush right qed marks, e.g. at end of proof
\usepackage{graphicx}
\usepackage{amsmath}
\usepackage{amssymb}
\usepackage{mathrsfs}
\usepackage{color}
\usepackage{bm}
\usepackage{ulem}
\DeclareMathOperator{\Tr}{Tr}
\newcommand{\ave}[1]{\ensuremath{\left\langle#1\right\rangle}}
\newcommand{\aves}[1]{\ensuremath{\langle#1\rangle}}

\journalname{Journal of Statistical Physics}
\begin{document}

\title{Universal Form of Stochastic Evolution for Slow Variables in Equilibrium Systems}
% \subtitle{Do you have a subtitle?\\ If so, write it here}
\titlerunning{Universal Form of Stochastic Evolution for Slow Variables} % if too long for running head

\author{Masato Itami \and Shin-ichi Sasa}

\institute{
M. Itami \at
Fukui Institute for Fundamental Chemistry, Kyoto University, Kyoto 606-8103, Japan \\
\email{itami@fukui.kyoto-u.ac.jp}
\and S. Sasa \at
Department of Physics, Kyoto University, Kyoto 606-8502, Japan \\
\email{sasa@scphys.kyoto-u.ac.jp}
}

\date{Received: date / Accepted: date}

\maketitle

\begin{abstract}
 % 100 - 150 wards
 Nonlinear, multiplicative Langevin equations for a complete set of slow variables in equilibrium systems are generally derived on the basis of the separation of time scales.
 The form of the equations is universal and equivalent to that obtained by Green.
 An equation with a nonlinear friction term for Brownian motion turns out to be an example of the general results.
 A key method in our derivation is to use different discretization schemes in a path integral formulation and the corresponding Langevin equation, which also leads to a consistent understanding of apparently different expressions for the path integral in previous studies.
% 4 - 6 keywords
\end{abstract}

\section{Introduction} \label{sec:intro}

% Onsager theory

We start with a brief review of Onsager theory \cite{OnsagerI,OnsagerII,Hashitsume,Onsager-Machlup,Machlup-Onsager,Groot-Mazur}. 
Let $\bm{X}=(X^i)_{i=1}^N$ be a complete set of unconstrained thermodynamic extensive variables of an isolated system.
Onsager theory formulates the deterministic dynamics of the thermodynamic variables in relaxation processes to the equilibrium state.
The time evolution is simply expressed as 
\begin{equation}
 \frac{\mathrm{d}X^i}{\mathrm{d}t}= \sum_{j} L^{ij} \frac{\partial S(\bm{X})}{\partial X^j},
  \label{onsager}
\end{equation}
where $S(\bm{X})$ is the thermodynamic entropy of the system, $\partial S(\bm{X})/ \partial X_j$ corresponds to the thermodynamic force, and $L^{ij}$ is called the Onsager coefficient.
The important consequence of Onsager theory is the reciprocity
\begin{equation}
 L^{ij}=L^{ji}.
  \label{recip}
\end{equation}
This nontrivial result was derived by studying fluctuations in equilibrium.
Concretely, the fluctuation is assumed to be described by a Langevin equation,
\begin{equation}
 \frac{\mathrm{d}X^i}{\mathrm{d}t}= \sum_{j} L^{ij} \frac{\partial S(\bm{X})}{\partial X^j} + \sum_{j} l^{ij}\xi^{j} ,
  \label{onsager-noise}
\end{equation}
with the Gaussian white noise satisfying $\ave{\xi^i(t) \xi^{j}(t')}=\delta(t-t')$.
This assumption means that the most probable regression process for a given fluctuation is equivalent to the relaxation dynamics, which is referred to as the regression hypothesis. 
Furthermore, according to equilibrium statistical mechanics, the stationary probability density of $\bm{X}$ is
\begin{equation}
 P_{\rm eq}(\bm{X})=\frac{1}{Z}\exp \left[ S(\bm{X})\right] ,
  \label{pss}
\end{equation}
where $Z$ is the normalization constant.
The time-reversibility of microscopic systems then provides the nontrivial relation 
\begin{equation}
 \sum_k l^{ik}l^{jk}=2L^{ij},
 \label{FDR}
\end{equation}
which leads to (\ref{recip}).
The relation (\ref{FDR}) is referred to as the fluctuation-dissipation relation of the second kind.

% nonlinear Onsager 

It should be noted that (\ref{onsager}) is a nonlinear equation for $\bm{X}$ because $S(\bm{X})$ is not necessarily a quadratic function.
In the argument above, $L^{ij}$ is assumed to be independent of $\bm{X}$.
Because dependence of $L^{ij}$ on $\bm{X}$ is expected in general cases, it is natural to consider generalized forms of (\ref{onsager}) and (\ref{onsager-noise}).
One approach assumes (\ref{onsager-noise}) with $L^{ij}(\bm{X})$ and $l^{ij}(\bm{X})$ as the starting equation, where a multiplication rule for $l^{ij}(\bm{X})$ and $\xi^j$ is specified.
Once a stochastic system is defined, the stationary distribution for $\bm{X}$ is determined.
We then find that the stationary distribution is not given by (\ref{pss}) for any multiplication rule for $l^{ij}(\bm{X})$ and $\xi^{j}$.
This means that there is no consistent description of (\ref{onsager-noise}), (\ref{pss}), and (\ref{FDR}) when dependence of $L^{ij}$ on $\bm{X}$ is considered. 
The important thing here is that a generalization of (\ref{onsager-noise}) with $L^{ij}(\bm{X})$ is not obvious at all.

% microscopic derivation 

We can now describe the dynamics of $\bm{X}$ on the basis of a Hamiltonian system consisting of atoms and molecules.
Suppose that $\bm{X}$ is a complete set of slow variables for the system.
Examples include a complete set of unconstrained extensive variables in thermodynamics.
Then, by using a separation of time scales, we may study the time evolution
of $\bm{X}$ from the microscopic mechanical description.
The result is that (\ref{onsager-noise}) is replaced by
\begin{equation}
  \frac{\mathrm{d}X^i}{\mathrm{d}t}= \mathcal{J}^{i}_{\mathrm{rev}} (\bm{X}) +\sum_{j} L^{ij}(\bm{X})\frac{\partial S(\bm{X})}{\partial X^{j}} +\sum_j \frac{\partial L^{ij}(\bm{X})}{\partial X^{j}} + \sum_{j} l^{ij}(\bm{X})\cdot \xi^{j},
   \label{onsager-noise-g}
\end{equation}
where the multiplication of $l^{ij}(\bm{X})$ and $\xi^{j}$ is interpreted as Ito-type, and $\mathcal{J}^{i}_{\mathrm{rev}}$ is the so-called reversible term that does not contribute to changes in entropy.

% history

There is a long history of studies on (\ref{onsager-noise-g}).
In~\cite{Green}, Green derived the Fokker--Planck equation corresponding to (\ref{onsager-noise-g}) by combining phenomenological arguments with microscopic considerations.
This was the genesis of (\ref{onsager-noise-g}).
After this paper, many further studies were performed.
For example, Green's derivation was improved in~\cite{Yamamoto} where fewer assumptions were used.
More formal studies under the microscopic description re-derived the Fokker--Planck equation using a projection-operator method~\cite{Zwanzig,ZwanzigETAL} and a nonequilibrium statistical operator method~\cite{Bashkirov-Zubarev}.
In another direction, the Fokker--Planck equation corresponding to (\ref{onsager-noise-g}) was also derived from a general Fokker--Planck equation by imposing a detailed balance condition~\cite{Graham-HakenI,Graham-HakenII}.
Finally, the Langevin equation (\ref{onsager-noise-g}) was derived directly from Liouville's equation using a nonlinear projection operator method~\cite{Kawasaki}.
Thus (\ref{onsager-noise-g}) was well established by 1975. 

% skewed  history - Graham and their followers

However, the result (\ref{onsager-noise-g}) is less well known nowadays.
There may be two reasons.
First, Graham attempted to develop a co-variant description of nonlinear Onsager theory.
Although this theory is complicated, many papers in this direction followed~\cite{Graham1977II,Grabert-Green,Grabert-Graham-Green,Hanggi,EyinkETAL}.
Unfortunately, we do not find a final answer in this direction, but more importantly, we consider such a generalization to not be necessary at all.
Equation (\ref{onsager-noise-g}) is sufficient to be general and universal.
Second, when $\bm{X}$ is a set of extensive variables in thermodynamics, the third term becomes higher-order than the second term from the estimates $S =O(\Omega)$ and $X^{i}= O(\Omega)$ for the system size $\Omega$.
Therefore, if one combines the system size expansion~\cite{Kampen,KampenBook} in deriving the equation for slow variables, the third term does not appear for thermodynamically normal systems.
Although this argument is correct, we emphasize that (\ref{onsager-noise-g}) can be a starting point for all systems, including small systems, once we identify a complete set of slow variables under equilibrium conditions.
That is, in our opinion, (\ref{onsager-noise-g}) should be recognized as a fundamental equation for slow variables. 

% microscopic derivation - our version

The main purpose of this paper is to re-derive (\ref{onsager-noise-g}) with particular emphasis of the separation of time scales and a universal asymptotic form of the probability density for time-averaged fluxes~\cite{Oono}.
We first assume a complete set of slow variables.
Let $\tau_{\rm macro}$ be the shortest time scale of the slow variables and $\tau_{\rm micro}$ be the largest time scale of the other dynamical variables.
Then, from the separation of time scales $\tau_{\rm micro} \ll \tau_{\rm macro}$, we can find $\Delta t$ such that $\tau_{\rm micro} \ll \Delta t \ll \tau_{\rm macro}$.
This $\Delta t $ plays two crucial roles in the derivation of the equation for slow variables.
First, because $\tau_{\rm micro} \ll \Delta t$, we can consider the central limit theorem for the time averaged flux as a universal form of the asymptotic behavior of the transition probability of the slow variables during a time interval $\Delta t$.
The time reversibility in microscopic Hamiltonian systems provides a restriction on the transition probability.
Second, because $\Delta t \ll \tau_{\rm macro}$, this universal form of the transition probability leads to the path integral form of a stochastic system.
This stochastic system is nothing but (\ref{onsager-noise-g}).
This concept is quite natural and general.
Indeed, one can interpret the arguments of Onsager and Green through this concept.
Nevertheless, as far as we know, there is no explicit presentation of the derivation of (\ref{onsager-noise-g}) with the universal asymptotic form of the probability density for time-averaged fluxes and the path integral formulation under $\tau_{\rm micro} \ll \Delta t \ll \tau_{\rm macro}$.
We thus expect that this paper will be instructive for understanding the universal form (\ref{onsager-noise-g}), and will also be useful for deriving the equation for slow variables even in systems out of equilibrium.

%% more explanation

Here, we point out the difference between our and previous approaches.
Our final goal is to establish a firm connection between a Langevin equation and a microscopic mechanical system.
The previous studies~\cite{Zwanzig,ZwanzigETAL,Bashkirov-Zubarev,Kawasaki} using a projection operator method or a nonequilibrium statistical operator method have the same motivation as ours.
Their methods use some physical approximation (such as a markovian approximation) just before obtaining a Langevin equation.
The validity of the approximation depends on observation time scales and details of a system, and their formulation is based on only an identity, which is useful but far from a physical principle.
Thus, their assumptions are out of scope of the theories.
Then, we aim to achieve our goal with physical principles.
From this motivation, we use the central limit theorem with the separation of time scales for connecting a microscopic mechanical and mesoscopic stochastic description in a mathematically and physically clear way.
This paper also differs from another type of derivation of the Langevin equation on the basis of arguments within stochastic processes~\cite{Graham-HakenI,Graham-HakenII}.
Their derivation is self-contained and elegant, but arguments relating to microscopic descriptions are out of scope of their theory.
As a technical remark, we note that they used the Kolmogorov forward and backward equations for restricting the form of the Langevin equation by imposing a detailed balance condition, while we directly use the transition probabilities.
Although we do not completely achieve our aim, we believe that it is important to show the outline of our approach even without a rigorous proof.
Our approach is not simply another derivation of known results, but provides a new direction of future studies.

% organization of this paper

The remainder of this paper is organized as follows.
In Section~\ref{sec:PImL}, as preliminaries for the argument, we review a path integral formulation for a discrete-time Langevin equation.
As described above, the central limit theorem for time-averaged fluxes is closely related to the path integral formulation of stochastic processes. 
The technical difficulty in the argument arises from its complicated expression, which may be entirely associated with the ill-defined nature of the multiplication of some quantities.  
To make the argument as clear as possible, we study a path integral form for discrete-time Langevin equations while keeping the time interval $dt$ finite.
We then find relations between different expressions of the path integral forms as first derived by Wissel \cite{Wissel}.
This recovers each of the correct but apparently different expressions for the path integral in~\cite{Lau-Lubensky,AronETAL,ChernyakETAL}.
In Section~\ref{sec:NLE_BP}, we consider as a special case a nonlinear Langevin equation for the momentum $\bm{P}$ of a Brownian particle of mass $M$ in a homogeneous environment of temperature $T$,
\begin{align}
 \frac{\mathrm{d}\bm{P}}{\mathrm{d}t} &= -\frac{\gamma(\bm{P})}{M}\bm{P} +\sqrt{2 \gamma(\bm{P}) T}\odot \bm{\xi}
 \notag
 \\
 &= -\frac{\gamma(\bm{P})}{M}\bm{P} + T\nabla \gamma (\bm{P}) + \sqrt{2 \gamma(\bm{P}) T}\cdot \bm{\xi},
  \label{non-lin-Lan}
\end{align}
where $\odot$ denotes multiplication with the anti-Ito rule.
Note that the Ito and anti-Ito rule will be explained in Subsection~\ref{subsec:PImL}.
Although (\ref{non-lin-Lan}) is well known, it has never been recognized as an example of (\ref{onsager-noise-g}) to the best of our knowledge.
Indeed, we can derive (\ref{non-lin-Lan}) by using the central limit theorem with the separation of time scales.
Then, in Section~\ref{sec:NLE_OT}, we derive the general formula (\ref{onsager-noise-g}).
%% notation
Throughout this paper, the Boltzmann constant $k_{\mathrm{B}}$ is set to unity.

%%%%%%%%%%%%%%%%%%%%%%%%%%%%%%%%%

\section{Preliminaries: Path Integral Formulation of Discrete Time Stochastic Systems} \label{sec:PImL}

\subsection{Model and Path Integral Formulation} \label{subsec:PImL}

Let $\bm{x}=(x^{1},\dots ,x^{N})$ be a collection of dynamical variables.
We study the time evolution of $\bm{x}$ for a fixed time interval $dt$.
We denote $x^{i}(ndt)$ by $x^{i}_{n}$, and we assume that $dx^{i}_{n}\equiv x^{i}_{n+1}-x^{i}_{n}$ satisfies
\begin{equation}
 d x^{i}_{n} = f^{i}(\bm{x}_{n}) dt + \sum_{j} g^{ij}(\bm{x}_{n}) dB^{j}_{n},
  \label{eq:L_multi}
\end{equation}
where $f^{i}$ and $g^{ij}$ are smooth functions of $\bm{x}$, $\bm{B}(t)$ is a standard $N$-dimensional Wiener process~\cite{Gardiner}, and $dB^{i}_{n} \equiv B^{i}(ndt + dt) - B^{i}(ndt)$ is a Gaussian white noise with mean zero and covariance $\aves{dB^{i}_{n}dB^{j}_{m}}=\delta^{ij}\delta_{nm}dt$.
Because the short time interval $dt$ can be considered to consist of shorter time intervals, we may use $dB^{i}_{n}dB^{j}_{m}=\delta^{ij}\delta_{nm}dt$ and ignore any $o(dt)$ terms in the Taylor expansion, which is known as It\^o's lemma.
Note that we obtain the It\^o stochastic differential equation from (\ref{eq:L_multi}) in the limit $dt \to 0$.

Denoting the probability density of finding $\bm{\chi}$ at time $t$ by
\begin{equation}
 P(\bm{\chi},t)=\aves{\delta(\bm{\chi}-\bm{x}_{t})} ,
\end{equation}
and using It\^o's lemma and (\ref{eq:L_multi}), we obtain the Fokker--Planck equation~\cite{Gardiner} 
\begin{align}
 \frac{\partial}{\partial t}P(\bm{\chi} ,t) &= - \sum_{i} \frac{\partial}{\partial \chi^{i}} \bigg[ f^{i}(\bm{\chi}) P(\bm{\chi} ,t)\bigg] + \sum_{i,j} \frac{\partial}{\partial \chi^{i}}\frac{\partial}{\partial \chi^{j}} \bigg[ G^{ij}(\bm{\chi}) P(\bm{\chi} ,t) \bigg]
 \label{eq:FP_a_multi}
\end{align}
in the limit $dt\to 0$ with
\begin{align}
 G^{ij}(\bm{x}) &= \frac{1}{2}\sum_{k} g^{ik}(\bm{x}) g^{jk}(\bm{x}).
\end{align}

We introduce a parameter $\alpha$ satisfying $0\leq \alpha\leq 1$, and define
\begin{align}
 \bar{x}_{n}^{i} &\equiv \alpha x_{n+1}^{i}+(1-\alpha ) x_{n}^{i}.
\end{align}
The purpose of this section is to express the transition probability $\mathcal{P}(\bm{x}_{n+1}\vert \bm{x}_{n})$ from $\bm{x}_{n}$ to $\bm{x}_{n+1}$ in terms of $d\bm{x}_{n}$ and $\bar{\bm{x}}_{n}$.
For any function $A(\bm{x})$ in the remainder of Section~\ref{sec:PImL}, we abbreviate $A(\bar{\bm{x}}_{n})$ and $\partial A(\bar{\bm{x}}_{n})/\partial \bar{x}^{i}_{n}$ to $A$ and $\partial^{i}A$, respectively.
We present the expression for $\mathcal{P}(\bm{x}_{n+1}\vert \bm{x}_{n})$ and derive it in the next subsection.
The transition probability is
\begin{align}
 \mathcal{P}(\bm{x}_{n+1}\vert \bm{x}_{n}) &= \frac{1}{\sqrt{(4\pi dt)^{N} \det \mathsf{G}}} \exp \Bigg[ -\frac{dt}{4}\sum_{i,j} \Delta^{i}_{n} (G^{-1})^{ij} \Delta^{j}_{n}
 \notag
 \\[3pt]
 & \qquad - \alpha \sum_{i} \partial^{i} f^{i} dt + \alpha^{2} \sum_{i,j} \partial^{i} \partial^{j}  G^{ij} dt\Bigg]
 \label{eq:PI_a_multi}
\end{align}
with
\begin{align}
 \Delta^{i}_{n} &\equiv \frac{dx^{i}_{n}}{dt} - f^{i}(\bar{\bm{x}}_{n}) + 2\alpha \sum_{k} \partial^{k} G^{ik}(\bar{\bm{x}}_{n}) ,
\end{align}
where $(G^{-1})^{ij}$ is the $ij$ component of the inverse of the matrix $\mathsf{G}=(G^{ij})$.
In~\cite{Wissel}, this expression for the transition probability was derived from the Fokker--Planck equation~(\ref{eq:FP_a_multi}).
Note that (\ref{eq:L_multi}) itself does not depend on $\alpha$.
In Appendix~\ref{sec:PIaL}, we check the normalization condition for the transition probability~(\ref{eq:PI_a_multi}) for $N=1$, and derive the Fokker--Planck equation~(\ref{eq:FP_a_multi}) from the transition probability~(\ref{eq:PI_a_multi}) in the limit $dt\to 0$.

Next, we consider the continuous-time limit of the transition probability.
To avoid divergence of the prefactor $1/\sqrt{(4\pi dt)^{N} \det \mathsf{G}}$ in the transition probability~(\ref{eq:PI_a_multi}), we rewrite (\ref{eq:PI_a_multi}) as
\begin{align}
 \mathcal{P}(\bm{x}_{n+1}\vert \bm{x}_{n}) = \int \frac{\mathrm{d}^{N}\bar{\bm{p}}_{n}}{(2\pi)^{N}}\; \exp \Big[ dt \mathcal{L}(\bm{x}_{n+1},\bar{\bm{p}}_{n}\vert \bm{x}_{n}) \Big]
 \label{eq:PI_con_multi}
\end{align}
with
\begin{align}
 \mathcal{L}(\bm{x}_{n+1},\bar{\bm{p}}_{n}\vert \bm{x}_{n}) &= - \sum_{i,j} \bar{p}^{i}_{n} G^{ij}(\bar{\bm{x}}_{n}) \bar{p}^{j}_{n} - i\sum_{i}\bar{p}^{i}_{n} \Delta^{i}_{n}
 \notag
 \\[2pt]
 & \qquad - \alpha \sum_{i} \partial^{i} f^{i}(\bar{\bm{x}}_{n}) + \alpha^{2}\sum_{i,j} \partial^{i} \partial^{j}  G^{ij}(\bar{\bm{x}}_{n}) ,
\end{align}
where $\bar{\bm{p}}_{n}=(\bar{p}^{1}_{n},\dots ,\bar{p}^{N}_{n})$ is interpreted as the conjugate momentum of $\bar{\bm{x}}_{n}$.
Using (\ref{eq:PI_con_multi}) repeatedly in each step and taking the limit $dt\to 0$, we can obtain the path integral for the Langevin equation.
Note that the Stratonovich convention ($\alpha=1/2$) in (\ref{eq:PI_con_multi}) may be convenient for a perturbative analysis based on the path integral formulation because the Stratonovich convention conserves the chain rule of differential calculus.

Finally, we compare (\ref{eq:PI_a_multi}) and (\ref{eq:PI_con_multi}) with some previous studies~\cite{Lau-Lubensky,AronETAL,ChernyakETAL}.
Instead of (\ref{eq:L_multi}), we consider
\begin{equation}
 d x^{i}_{n} = \tilde{F}^{i}(\tilde{\bm{x}}_{n}) dt + \sum_{j} g^{ij}(\tilde{\bm{x}}_{n}) dB^{j}_{n}
  \label{eq:L_multi_at}
\end{equation}
with
\begin{equation}
 \tilde{x}_{n}^{i} \equiv \tilde{\alpha} x_{n+1}^{i}+(1-\tilde{\alpha} ) x_{n}^{i},
\end{equation}
where $\tilde{F}^{i}$ is some smooth function, and $\tilde{\alpha}$ is a parameter satisfying $0\leq \tilde{\alpha} \leq 1$.
Here, $\tilde{\alpha}=0$, $\tilde{\alpha}=1/2$, and $\tilde{\alpha}=1$ correspond to the It\^o, Stratonovich, and anti-It\^o convention, respectively, in the limit $dt \to 0$.
Using It\^o's lemma, we can rewrite (\ref{eq:L_multi_at}) as
\begin{equation}
 d x^{i}_{n} = \tilde{F}^{i}(\bm{x}_{n}) dt + \tilde{\alpha} \sum_{j,k} g^{kj}(\bm{x}_{n}) \frac{\partial g^{ij}(\bm{x}_{n})}{\partial x^{k}_{n}}dt + \sum_{j} g^{ij}(\bm{x}_{n}) dB^{j}_{n}.
 \label{eq:L_multi_at_ito}
\end{equation}
Thus, we can obtain the transition probability $\mathcal{P}(\bm{x}_{n+1}\vert \bm{x}_{n})$ for (\ref{eq:L_multi_at}) by using (\ref{eq:PI_a_multi}) with
\begin{equation}
 f^{i}=\tilde{F}^{i}+\tilde{\alpha}\sum_{j,k}g^{kj}\partial^{k}g^{ij}.
  \label{eq:f_and_F}
\end{equation}
Note that $\tilde{\alpha}$ may be different from $\alpha$.
When $\alpha=\tilde{\alpha}$, (\ref{eq:PI_con_multi}) with (\ref{eq:f_and_F}) is equivalent to the results given in~\cite{Lau-Lubensky,AronETAL}.
When $(\alpha,\tilde{\alpha})=(1/2,1)$, (\ref{eq:PI_a_multi}) with (\ref{eq:f_and_F}) is equivalent to the results given in~\cite{ChernyakETAL}.

\subsection{Derivation}

We derive the transition probability~(\ref{eq:PI_a_multi}) from (\ref{eq:L_multi}) without using the Fokker--Planck equation (\ref{eq:FP_a_multi}).
Using It\^o's lemma, we first rewrite (\ref{eq:L_multi}) in terms of $\bar{\bm{x}}_{n}$ as
\begin{equation}
 d x^{i}_{n} = F^{i} dt + \sum_{j} g^{ij} dB^{j}_{n}
  \label{eq:L_a_multi}
\end{equation}
with
\begin{equation}
 F^{i} = f^{i} - \alpha \sum_{j,k} g^{kj} \partial^{k} g^{ij}.
  \label{eq:Ff_rela}
\end{equation}
Because $dB^{i}_{n}$ is the Gaussian white noise with covariance $\aves{dB^{i}_{n}dB^{j}_{m}}=\delta^{ij}\delta_{nm}dt$, the probability density of $\{ dB^{i}_{n}\}$ is given by
\begin{equation}
 P(\{ dB^{i}_{n}\}) = \frac{1}{(\sqrt{2\pi dt})^{N}} \exp \left[ -\frac{1}{2dt}\sum_{i} (dB^{i}_{n})^{2}\right] .
  \label{eq:P_Bn_multi}
\end{equation}
Using (\ref{eq:L_multi}), $\bm{x}_{n+1}$ is uniquely determined by $\bm{x}_{n}$ and $\{ dB^{i}_{n}\}$.
Thus, we have
\begin{equation}
 \mathcal{P}(\bm{x}_{n+1}\vert \bm{x}_{n}) = P(\{ dB^{i}_{n}\}) \vert \det \mathcal{J} \vert ,
  \label{eq:trans_XB}
\end{equation}
where $\mathcal{J}=(J^{ij})$ is the Jacobian matrix defined by
\begin{equation}
 J^{ij} \equiv \frac{\partial (dB^{i}_{n})}{\partial x^{j}_{n+1}}.
\end{equation}

We next calculate the determinant of the Jacobian matrix $\mathcal{J}$.
Differentiating both sides of (\ref{eq:L_a_multi}) with respect to $x^{l}_{n+1}$, we obtain
\begin{equation}
 \sum_{j} g^{ij} J^{jl} = \mu^{il},
  \label{eq:diff_L_multi}
\end{equation}
where the matrix $\mathcal{M}=(\mu^{il})$ is given by
\begin{equation}
 \mu^{il} = \delta^{il} - \alpha \partial^{l}F^{i}dt - \alpha \sum_{j} \partial^{l}g^{ij}dB^{j}_{n}.
\end{equation}
Denoting the identity matrix of size $N$ by $I_{N}$, we define the matrix $\widetilde{\mathcal{M}}=(\tilde{\mu}^{ij})$ by $\widetilde{\mathcal{M}}\equiv I_{N}-\mathcal{M}$.
Using It\^o's lemma, the determinant of the matrix $\mathcal{M}$ is
\begin{align}
 \det \mathcal{M} &= \det \left[ \exp \left( \log \mathcal{M}\right)\right]
 \notag
 \\[3pt]
 &= \exp \left[ \Tr \left( \log \mathcal{M}\right) \right]
 \notag
 \\[3pt]
 &= \exp \left[ -\Tr \widetilde{\mathcal{M}}-\frac{1}{2}\Tr \widetilde{\mathcal{M}}^{2}\right]
 \notag
 \\[3pt]
 &= \exp \Bigg[ - \alpha \sum_{i} \partial^{i} F^{i}dt - \alpha \sum_{i,j} \partial^{i}g^{ij}dB^{j}_{n}  - \frac{\alpha^{2}}{2} \sum_{i,j,k} \partial^{i}g^{jk}\partial^{j}g^{ik} dt \Bigg] .
\end{align}
Because (\ref{eq:diff_L_multi}) leads to
\begin{equation}
 \det \mathcal{G} \det \mathcal{J} = \det \mathcal{M}
\end{equation}
with $\mathcal{G}=(g^{ij})$, we obtain
\begin{align}
 \det \mathcal{J} &= \frac{1}{\det \mathcal{G}} \exp \Bigg[ - \alpha \sum_{i} \partial^{i} F^{i}dt - \alpha \sum_{i,j} \partial^{i}g^{ij}dB^{j}_{n} - \frac{\alpha^{2}}{2} \sum_{i,j,k} \partial^{i}g^{jk}\partial^{j}g^{ik} dt \Bigg] .
 \label{eq:Jacobian_multi}
\end{align}

Substituting (\ref{eq:P_Bn_multi}) and (\ref{eq:Jacobian_multi}) into (\ref{eq:trans_XB}), we obtain
\begin{align}
 \mathcal{P}(\bm{x}_{n+1}\vert \bm{x}_{n}) &= \frac{1}{(\sqrt{2\pi dt})^{N}\vert \det \mathcal{G}\vert} \exp \Bigg[ -\frac{1}{2dt}\sum_{i}\bigg[ dB^{i}_{n}+\alpha \sum_{k}\partial^{k} g^{ki}dt\bigg]^{2}
 \notag
 \\[3pt]
 & \qquad - \alpha \sum_{i} \partial^{i} F^{i}dt + \frac{\alpha^{2}}{2} \sum_{i,j,k} \left[ \partial^{i}g^{ik}\partial^{j}g^{jk} - \partial^{i}g^{jk}\partial^{j}g^{ik}\right] dt\Bigg]
 \notag
 \\[3pt]
 &= \frac{1}{(\sqrt{2\pi dt})^{N}\vert \det \mathcal{G}\vert} \exp \Bigg[ -\frac{dt}{2}\sum_{i} \bigg[ \sum_{j} (g^{-1})^{ij}
 \notag
 \\[3pt]
 & \qquad \bigg( \frac{dx^{j}_{n}}{dt}-F^{j} + \alpha \sum_{k,l} g^{jl}\partial^{k}g^{kl} \bigg) \bigg]^{2} - \alpha \sum_{i} \partial^{i} F^{i} dt 
 \notag
 \\[3pt]
 & \qquad + \frac{\alpha^{2}}{2} \sum_{i,j,k} \left[ \partial^{i}g^{ik}\partial^{j}g^{jk} - \partial^{i}g^{jk}\partial^{j}g^{ik}\right] dt\Bigg] ,
 \label{eq:PI_a_at_multi}
\end{align}
where $(g^{-1})^{ij}$ is the $ij$ component of the inverse of the matrix $\mathcal{G}$.
Using (\ref{eq:Ff_rela}), $\det \mathsf{G} = (\det \mathcal{G})^{2}/2^{N}$, and
\begin{equation}
 g^{ik}\partial^{j} g^{jk} - g^{jk}\partial^{j} g^{ik} = \partial^{j} (g^{ik}g^{jk}) -2g^{jk}\partial^{j}g^{ik},
\end{equation}
we can rewrite (\ref{eq:PI_a_at_multi}) as (\ref{eq:PI_a_multi}).

\section{Example: Brownian Motion} \label{sec:NLE_BP}

\subsection{Setup}

In this section, we study the motion of a single Brownian particle in a fluid (heat bath).
We derive (\ref{non-lin-Lan}) under the assumption that the relaxation time of the momentum $\tau_{\mathrm{macro}}$ is much larger than the time scales of the other degrees of freedom.
We describe the system as a Hamiltonian system.
The system consists of $N$ bath particles of mass $m$ and a Brownian particle of mass $M$ in a cube of side length $L$.
For simplicity, periodic boundary conditions are assumed.
Let $(\bm{r}_{i}, \bm{p}_{i})$ $(1\leq i\leq N)$ be the position and momentum of the $i$th bath particle, and $(\bm{R}, \bm{P})$ be those of the Brownian particle.
The collection of the positions and momenta of all particles is denoted by $\Gamma = (\bm{r}_{1},\bm{p}_{1},\dots ,\bm{r}_{N},\bm{p}_{N},\bm{R},\bm{P})$, which represents the microscopic state of the system.
For any state $\Gamma$, we denote its time reversal by $\Gamma^{\ast}$, namely, the state obtained by reversing all the momenta, and denote the time reversal of $\bm{P}$ by $\bm{P}^{\ast}=-\bm{P}$.
For convenience, we denote the microscopic state excluding the momentum of the Brownian particle by $\tilde{\Gamma} = (\bm{r}_{1},\bm{p}_{1},\dots ,\bm{r}_{N},\bm{p}_{N},\bm{R})$.

The Hamiltonian of the system is given by 
\begin{equation}
 H(\Gamma) =  \tilde{H}(\tilde{\Gamma}) + \frac{\vert\bm{P}\vert^2}{2M}
  \label{eq:Hamiltonian}
\end{equation}
with
\begin{equation}
 \tilde{H}(\tilde{\Gamma}) = \sum_{i=1}^{N} \Bigg[ \frac{\vert \bm{p}_{i}\vert^{2}}{2m} + \sum_{j>i} \Phi_{\rm int}(\vert\bm{r}_{i}-\bm{r}_{j}\vert ) + \Phi_{\rm B}(\vert \bm{r}_{i}-\bm{R}\vert )\Bigg] ,
\end{equation}
where $\Phi_{\rm int}$ is a short-range interaction potential between two bath particles, and $\Phi_{\rm B}$ is that between a bath particle and the Brownian particle.
Then the Hamiltonian satisfies the time-reversal symmetry
\begin{equation}
 H(\Gamma^{\ast}) = H(\Gamma).
  \label{eq:time_reversal_symmetry}
\end{equation}
For the Hamiltonian equations with a given state $\Gamma$ at $t=0$, $\Gamma_{t}$ denotes the solution at time $t$.
In this setup, energy is conserved, that is,
\begin{equation}
 H(\Gamma_{t})=H(\Gamma),
  \label{eq:energy_conservation}
\end{equation}
and Liouville's theorem that 
\begin{equation}
 \left\vert \frac{\partial \Gamma_{t}}{\partial \Gamma }\right\vert = 1
  \label{eq:Liouville}
\end{equation}
holds.
The total force acting on the Brownian particle is given by
\begin{align}
 \bm{F}(\Gamma) &= -\frac{\partial H(\Gamma)}{\partial \bm{R}}
 \notag
 \\
 &= - \sum_{i=1}^{N} \frac{\partial \Phi_{\rm B}(\vert \bm{r}_{i}-\bm{R}\vert )}{\partial \bm{R}}.
 \label{eq:def_force}
\end{align}
For convenience, we abbreviate $A(\Gamma_{t})$ to $A_{t}$ for any function $A$.
The equation of motion for the Brownian particle is
\begin{equation}
 \frac{\mathrm{d}\bm{P}_{t}}{\mathrm{d}t}=\bm{F}_{t}.
  \label{eq:EOM}
\end{equation}

We assume that the system in equilibrium is at temperature $T$.
Suppose that the momentum of the Brownian particle is $\bm{P}_{\mathrm{i}}$ at an initial time.
Then the other mechanical state $\tilde{\Gamma}$ is sampled according to the probability density
\begin{align}
 \tilde{P}_{\mathrm{eq}}(\tilde{\Gamma}) = \exp \left[ -\frac{\tilde{H}(\tilde{\Gamma}) -\tilde{\Psi}_{\rm eq}}{T}\right] ,
\end{align}
where $\tilde{\Psi}_{\rm eq}$ is the normalization constant.
The Hamiltonian equation determines the value of $\bm{P}$ at time $t$.
By taking the average over initial realizations of $\tilde{\Gamma}$, we determine the probability density of $\bm{P}=\bm{P}_{\mathrm{f}}$ at time $t$ for a given $\bm{P}_{\mathrm{i}}$ at time $0$ in the form
\begin{align}
 \mathcal{P}_{t}(\bm{P}_{\mathrm{f}}\vert\bm{P}_{\mathrm{i}}) \equiv \int \mathrm{d}\Gamma \; \tilde{P}_{\mathrm{eq}}(\tilde{\Gamma}) \delta (\bm{P}-\bm{P}_{\mathrm{i}}) \delta (\bm{P}_{t}-\bm{P}_{\mathrm{f}}) .
 \label{eq:TP}
\end{align}
It should be noted that
\begin{equation}
 \int \mathrm{d}^{3} \bm{P}_{\mathrm{f}}\; \mathcal{P}_{t}(\bm{P}_{\mathrm{f}}\vert\bm{P}_{\mathrm{i}}) = 1.
\end{equation}
When we describe the motion of the Brownian particle, the position $\bm{R}$ should be treated in the same manner as $\bm{P}$ because $\mathrm{d}\bm{R}/\mathrm{d}t=\bm{P}$.
For simplicity, we consider space translational symmetric systems, so that we do not need to specify the position at the initial time.
If one considers an external potential acting on the Brownian particle, then $\mathcal{P}_{t}(\bm{P}_{\mathrm{f}}\vert\bm{P}_{\mathrm{i}})$ should be replaced by $\mathcal{P}_{t}(\bm{P}_{\mathrm{f}},\bm{R}_{\mathrm{f}}\vert\bm{P}_{\mathrm{i}},\bm{R}_{\mathrm{i}})$.

Here, the most important property of $\mathcal{P}_{t}$ is
\begin{equation}
 \mathcal{P}_{t}(\bm{P}_{\mathrm{f}}\vert\bm{P}_{\mathrm{i}}) P_{\mathrm{MB}}(\bm{P}_{\mathrm{i}}) = \mathcal{P}_{t}(\bm{P}_{\mathrm{i}}^{\ast}\vert\bm{P}_{\mathrm{f}}^{\ast}) P_{\mathrm{MB}}(\bm{P}_{\mathrm{f}}) ,
  \label{eq:DB}
\end{equation}
where we denote the Maxwell--Boltzmann distribution by
\begin{equation}
 P_{\mathrm{MB}}(\bm{P}) = (2\pi MT)^{-3/2}\exp \left[ -\frac{\vert\bm{P}\vert^{2}}{2MT}\right] .
  \label{eq:MBdist}
\end{equation}
Property~(\ref{eq:DB}) is called the detailed balance condition.
Using microscopic reversibility $(\Gamma^{\ast})_{-t}=(\Gamma_{t})^{\ast}$, (\ref{eq:time_reversal_symmetry}), (\ref{eq:energy_conservation}), and (\ref{eq:Liouville}), we can obtain the detailed balance condition~(\ref{eq:DB}).

\subsection{Assumptions}

Let $\tau_{\mathrm{micro}}$ be the correlation time of the force acting on the Brownian particle, and $\tau_{\mathrm{macro}}$ be the relaxation time of the momentum of the Brownian particle.
We have the separation of time scales represented by $\tau_{\mathrm{micro}}\ll \tau_{\mathrm{macro}}$ because we assume that the relaxation time of the momentum is much larger than the time scales of the other degrees of freedom.

We define the time-averaged total force acting on the Brownian particle as
\begin{equation}
 \bar{\bm{F}}(\Gamma) \equiv \frac{1}{\Delta t}\int_{0}^{\Delta t} \mathrm{d}s\; \bm{F}(\Gamma_{s}),
  \label{eq:def_aveF}
\end{equation}
where $\Delta t$ is a finite time interval that satisfies $\tau_{\mathrm{micro}}\ll \Delta t \ll \tau_{\mathrm{macro}}$.
We define the conditional probability density of $\bar{\bm{F}}$ given $\bm{P}_{\mathrm{i}}$ by
\begin{align}
 \widetilde{\mathcal{P}}(\bar{\bm{F}}\vert \bm{P}_{\mathrm{i}}) \equiv \int \mathrm{d}\Gamma \; \tilde{P}_{\mathrm{eq}}(\tilde{\Gamma}) \delta (\bm{P}-\bm{P}_{\mathrm{i}}) \delta (\bar{\bm{F}}(\Gamma)-\bar{\bm{F}}).
\end{align}
Considering $\Delta t \gg \tau_{\mathrm{micro}}$, we may employ the central limit theorem, according to which and the isotropic property in equilibrium we have the following Gaussian form of $\widetilde{\mathcal{P}}(\bar{\bm{F}}\vert \bm{P}_{\mathrm{i}})$:
\begin{align}
 \widetilde{\mathcal{P}}(\bar{\bm{F}}\vert \bm{P}_{\mathrm{i}}) = C \exp \left[ -\Delta t\, \mathcal{I}(\bar{\bm{F}}\vert \bm{P}_{\mathrm{i}})\right]
 \label{eq:LDP_F}
\end{align}
with
\begin{equation}
 \mathcal{I}(\bar{\bm{F}}\vert \bm{P}_{\mathrm{i}}) = \frac{\vert \bm{\bar{F}} - \bm{\mathcal{F}} (\bm{P}_{\mathrm{i}})\vert^{2}}{4T\gamma (\bm{P}_{\mathrm{i}})},
  \label{eq:CLT}
\end{equation}
where $C$ is the normalization constant given by
\begin{equation}
 C = \left[ \frac{\Delta t}{4\pi T \gamma (\bm{P}_{\mathrm{i}})}\right]^{3/2},
\end{equation}
$\bm{\mathcal{F}} (\bm{P}_{\mathrm{i}})$ is the most probable value of $\bar{\bm{F}}$, and $2T\gamma(\bm{P})$ is the dispersion of fluctuations of $\bar{\bm{F}}$.
The dispersion $\gamma (\bm{P})$ is assumed to be positive for any $\bm{P}$.
Because of the space-reflection symmetry, $\bm{\mathcal{F}}$ and $\gamma$ satisfy
\begin{align}
 \bm{\mathcal{F}}(\bm{P}^{\ast}) &= - \bm{\mathcal{F}}(\bm{P}),
 \\[3pt]
 \gamma (\bm{P}^{\ast}) &= \gamma (\bm{P}) ,
\end{align}
respectively.
Assumption (\ref{eq:LDP_F}) with (\ref{eq:CLT}) is essential for our derivation of a nonlinear Langevin equation for the Brownian particle.
Note that we may prove this assumption when bath particles collisions with a Brownian particle can be regarded as independent.

\subsection{Derivation}

Using the equation of motion (\ref{eq:EOM}), $\bar{\bm{F}}(\Gamma)$ defined in (\ref{eq:def_aveF}) can be rewritten as
\begin{equation}
 \bar{\bm{F}}(\Gamma) = \frac{\bm{P}_{\Delta t}-\bm{P}}{\Delta t}.
\end{equation}
Thus, by changing variables from $\bar{\bm{F}}(\Gamma)$ to $\bm{P}_{\Delta t}$ in (\ref{eq:LDP_F}) with (\ref{eq:CLT}), we obtain
\begin{align}
 \mathcal{P}_{\Delta t}(\bm{P}_{\mathrm{f}}\vert\bm{P}_{\mathrm{i}}) &= \widetilde{\mathcal{P}}(\bar{\bm{F}}\vert \bm{P}_{\mathrm{i}}) / (\Delta t)^{3}
 \notag
 \\[3pt]
 & = \left[ 4\pi \Delta t T \gamma (\bm{P}_{\mathrm{i}})\right]^{-3/2} \exp \left[ -\frac{\Delta t}{4T\gamma (\bm{P}_{\mathrm{i}})} \left\vert \frac{\bm{P}_{\mathrm{f}}-\bm{P}_{\mathrm{i}}}{\Delta t} - \bm{\mathcal{F}} (\bm{P}_{\mathrm{i}}) \right\vert^{2}\right] .
 \label{eq:TP_ito}
\end{align}
When we compare (\ref{eq:TP_ito}) with (\ref{eq:PI_a_multi}) for $\alpha =0$, we can describe the discrete time evolution of $\bm{P}$ as the discrete stochastic system (\ref{eq:L_multi}).
Now, taking the limit $\Delta t/\tau_{\mathrm{macro}}\to 0$, we obtain the Langevin equation
\begin{align}
 \frac{\mathrm{d}\bm{P}_{t}}{\mathrm{d}t} = \bm{\mathcal{F}}(\bm{P}_{t}) + \sqrt{2T\gamma (\bm{P}_{t})} \cdot \bm{\xi}_{t} ,
 \label{eq:NLE_ito}
\end{align}
where $\bm{\xi}_{t}$ is the zero-mean Gaussian white noise with covariance $\langle \xi^{a}_{t}\xi^{b}_{s}\rangle = \delta^{ab} \delta(t-s)$ and $\cdot$ denotes multiplication with the It\^o rule.

Next, we express $\bm{\mathcal{F}}$ in terms of $\gamma$ from the detailed balance condition (\ref{eq:DB}).
Using (\ref{eq:PI_a_multi}) with $\alpha=1/2$, we can rewrite the transition probability~(\ref{eq:TP_ito}) in terms of $\bm{P}_{\mathrm{m}}\equiv (\bm{P}_{\mathrm{f}}+\bm{P}_{\mathrm{i}})/2$ as
\begin{align}
 \mathcal{P}_{\Delta t}(\bm{P}_{\mathrm{f}}\vert\bm{P}_{\mathrm{i}}) & = \left[ 4\pi \Delta t T \gamma (\bm{P}_{\mathrm{m}})\right]^{-3/2} \exp \Bigg[ -\frac{\Delta t}{4T\gamma (\bm{P}_{\mathrm{m}})} \left\vert \frac{\bm{P}_{\mathrm{f}}-\bm{P}_{\mathrm{i}}}{\Delta t} - \widetilde{\bm{\mathcal{F}}} (\bm{P}_{\mathrm{m}}) \right\vert^{2}
 \notag
 \\[3pt]
 & \qquad -\frac{\Delta t}{2}\sum_{a=x,y,z}\frac{\partial \mathcal{F}^{a}(\bm{P}_{\mathrm{m}})}{\partial P^{a}_{\mathrm{m}}} +\frac{\Delta t}{4}\sum_{a=x,y,z} T\frac{\partial^{2}\gamma (\bm{P}_{\mathrm{m}})}{\partial P^{a}_{\mathrm{m}} \partial P^{a}_{\mathrm{m}}}\Bigg] ,
 \label{eq:PI_str}
\end{align}
with
\begin{align}
 \widetilde{\mathcal{F}}^{a} (\bm{P}_{\mathrm{m}}) &\equiv \mathcal{F}^{a} (\bm{P}_{\mathrm{m}}) -T\frac{\partial \gamma (\bm{P}_{\mathrm{m}})}{\partial P^{a}_{\mathrm{m}}} ,
 \label{eq:def_Ftilde}
\end{align}
where the superscript $a$ represents the indices in Cartesian coordinates $(x,y,z)$, and we have ignored all $o(\Delta t)$ terms because these terms are irrelevant in the limit $\Delta t/\tau_{\mathrm{macro}}\to 0$.
Note that the Stratonovich convention ($\alpha=1/2$) is convenient when using (\ref{eq:DB}) because it has the property that the forward and backward paths are evaluated at the same points~\cite{Lau-Lubensky}.
Substituting (\ref{eq:PI_str}) into (\ref{eq:DB}) with (\ref{eq:MBdist}), we obtain
\begin{align}
 \widetilde{\bm{\mathcal{F}}} (\bm{P}_{\mathrm{m}}) &= - \frac{\gamma (\bm{P}_{\mathrm{m}})}{M}\bm{P}_{\mathrm{m}},
 \label{eq:FDT}
\end{align}
which is called the fluctuation dissipation relation of the second kind.
From (\ref{eq:NLE_ito}), (\ref{eq:def_Ftilde}), and (\ref{eq:FDT}), we have
\begin{align}
 \frac{\mathrm{d}P_{t}^{a}}{\mathrm{d}t} &= - \frac{\gamma (\bm{P}_{t})}{M}P_{t}^{a} + T\frac{\partial \gamma (\bm{P}_{t})}{\partial P^{a}_{t}} + \sqrt{2T\gamma (\bm{P}_{t})} \cdot \xi_{t}^{a}
 \notag
 \\
 &= - \frac{\gamma (\bm{P}_{t})}{M}P_{t}^{a} + \sqrt{2T\gamma (\bm{P}_{t})} \odot \xi_{t}^{a},
 \label{eq:NLE_str}
\end{align}
which is equivalent to (\ref{non-lin-Lan}).
From (\ref{eq:NLE_str}), $\gamma (\bm{P}_{t})$ can be interpreted as a nonlinear friction coefficient.

\section{Generalization} \label{sec:NLE_OT}

\subsection{Motivation}

We consider fluctuations of a system in equilibrium. 
There is a special set of variables whose time scales are well separated from those of the other dynamical degrees of freedom.
We refer to such a set as a complete  set of slow variables and denote it by $\bm{X}=(X^1,X^2,\dots, X^N)$.
For the example in the previous section, $\bm{X}=(\bm{R},\bm{P})$.
As a different example, one may consider fluctuations in a thermodynamically isolated system separated into two regions by a freely movable diabatic wall.
In this case, unconstrained thermodynamic extensive variables, the energy and the volume in one region are assumed to form a complete set of slow variables.
Furthermore, hydrodynamic fluctuations, which are long-wavelength fluctuations of locally conserved quantities, in an equilibrium liquid are another example of a complete set of slow variables.
For simplicity, we assume that the Hamiltonian of the microscopic mechanical system is symmetric with respect to the time-reversal operation.
For such a system, the probability density of $\bm{X}$ is denoted by
\begin{equation}
 P_{\mathrm{eq}}(\bm{X}) = \frac{1}{Z}\exp \left[ S(\bm{X})\right] ,
\label{eq:dist_m}
\end{equation}
where $Z$ is the normalization constant, and $S(\bm{X}^{\ast})=S(\bm{X})$ for the time reversal $\bm{X}^{\ast}$ of $\bm{X}$.
The physical interpretation of $S$ depends on the system being studied.
For example, $S(\bm{X})$ corresponds to entropy when thermodynamic fluctuations in an isolated system are considered.
For other cases, $S(\bm{X})$ should be read from the form of the stationary distribution.
Suppose that the system is in equilibrium.
We then expect that the time evolution of $\bm{X}$ can be described by a Langevin equation.
In this section, we derive the equation by generalizing the arguments in the previous section.

\subsection{Basic concept}

On the basis of a microscopic mechanical description, we can define the conditional probability density of $\bm{X}=\bm{X}_{\mathrm{f}}$ at time $t$, denoted by $\mathcal{P}_{t}(\bm{X}_{\mathrm{f}} \vert \bm{X}_{\mathrm{i}})$, provided that $\bm{X}=\bm{X}_{\mathrm{i}}$ at time $0$.
There are two important properties of this probability density.
First, following the central limit theorem, we assume the Gaussian form of form of the probability density for the time averaged flux written as $(\bm{X}_{\Delta t}-\bm{X} )/\Delta t$.
The result (\ref{eq:TP_ito}) in the previous section becomes 
\begin{align}
 \mathcal{P}_{\Delta t}(\bm{X}_{\mathrm{f}}\vert\bm{X}_{\mathrm{i}}) & = \frac{1}{\sqrt{(4\pi \Delta t)^{N} \det \mathsf{L} (\bm{X}_{\mathrm{i}})}} \exp \Bigg[ -\frac{\Delta t}{4} \sum_{i,j} (L^{-1})^{ij}(\bm{X}_{\mathrm{i}})
 \notag
 \\[3pt]
 & \qquad \times \bigg[ \frac{X^{i}_{\mathrm{f}}-X^{i}_{\mathrm{i}}}{\Delta t}-\mathcal{J}^{i}(\bm{X}_{\mathrm{i}})\bigg] \bigg[ \frac{X^{j}_{\mathrm{f}}-X^{j}_{\mathrm{i}}}{\Delta t} -\mathcal{J}^{j}(\bm{X}_{\mathrm{i}})\bigg] \Bigg] .
 \label{eq:TP_ito_m}
\end{align}
where $\mathcal{J}^{i}$ is the most probable value of the time averaged flux, and $2L^{ij}$ is the dispersion matrix.
We assume that the matrix $\mathsf{L} = (L^{ij})$ is positive definite.
This means that each $X^{i}_{\mathrm{f}}$ is not uniquely determined by $\bm{X}_{\mathrm{i}}$.
If $X^{i}_{\mathrm{f}}$ is uniquely determined by $\bm{X}_{\mathrm{i}}$, such as $X^{i}_{\mathrm{f}}=X^{i}_{\mathrm{i}}+\mathcal{J}^{i}(\bm{X}_{\mathrm{i}})\Delta t$, then we multiply the right-hand side of (\ref{eq:TP_ito_m}) by $\delta (X^{i}_{\mathrm{f}}-X^{i}_{\mathrm{i}}-\mathcal{J}^{i}(\bm{X}_{\mathrm{i}})\Delta t)$ and consider the submatrix formed by deleting the $i$th row and $i$th column of $\mathsf{L}$.
Second, from the reversibility of microscopic Hamiltonian systems, we can obtain
\begin{equation}
 \mathcal{P}_{t}(\bm{X}_{\mathrm{f}}\vert\bm{X}_{\mathrm{i}}) P_{\mathrm{eq}}(\bm{X}_{\mathrm{i}}) = \mathcal{P}_{t}(\bm{X}_{\mathrm{i}}^{\ast}\vert\bm{X}_{\mathrm{f}}^{\ast}) P_{\mathrm{eq}}(\bm{X}_{\mathrm{f}}).
  \label{eq:DB_m}
\end{equation}
Then, by substituting (\ref{eq:TP_ito_m}) into  (\ref{eq:DB_m}), we obtain a possible form of $\mathcal{J}^{i}(\bm{X})$ and a symmetry property of $\mathsf{L}$.
For convenience, we denote $\bm{X}^{*}$ by $\bm{\epsilon}\bm{X}=(\epsilon^{1}X^{1},\epsilon^{2}X^{2},\dots ,\epsilon^{N}X^{N})$, where $\epsilon^{i}=+1$ or $-1$ for $X^{i}$.
We decompose $\mathcal{J}^{i}$ into two parts,
\begin{equation}
 \mathcal{J}^{i}(\bm{X}) = \mathcal{J}^{i}_{\mathrm{rev}}(\bm{X}) + \mathcal{J}^{i}_{\mathrm{irr}}(\bm{X})
\end{equation}
with
\begin{align}
 \mathcal{J}^{i}_{\mathrm{rev}}(\bm{X}) &\equiv \frac{\mathcal{J}^{i}(\bm{X})-\epsilon^{i}\mathcal{J}^{i}(\bm{X}^{*})}{2}, \\[5pt]
 \mathcal{J}^{i}_{\mathrm{irr}}(\bm{X}) &\equiv \frac{\mathcal{J}^{i}(\bm{X})+\epsilon^{i}\mathcal{J}^{i}(\bm{X}^{*})}{2},
\end{align}
which satisfy $\mathcal{J}^{i}_{\mathrm{rev}}(\bm{X}^{*}) = -\epsilon^{i}\mathcal{J}^{i}_{\mathrm{rev}}(\bm{X})$ and $\mathcal{J}^{i}_{\mathrm{irr}}(\bm{X}^{*}) = \epsilon^{i}\mathcal{J}^{i}_{\mathrm{irr}}(\bm{X})$.
We also define the matrix $\mathsf{L}_{\mathrm{T}}=(L^{ij}_{\mathrm{T}})$ by
\begin{equation}
 L^{ij}_{\mathrm{T}}(\bm{X}) = \epsilon^{i}\epsilon^{j}L^{ij}(\bm{X}^{*}).
\end{equation}
Note that $\det \mathsf{L}_{\mathrm{T}}(\bm{X})=\det \mathsf{L}(\bm{X}^{*})$.

\subsection{Result}

% determination of the force 

Direct substitution of (\ref{eq:TP_ito_m}) into (\ref{eq:DB_m}) would result in a complicated form, so we use a trick.
Considering that (\ref{eq:PI_a_multi}) holds for any $\alpha$, as in the previous section, we can rewrite (\ref{eq:TP_ito_m}) by changing $\alpha=0$ to $\alpha=1/2$.
The result is
\begin{align}
 \mathcal{P}_{\Delta t}(\bm{X}_{\mathrm{f}}\vert\bm{X}_{\mathrm{i}}) & = \frac{1}{\sqrt{(4\pi \Delta t)^{N} \det \mathsf{L} (\bm{X}_{\mathrm{m}})}} \exp \Bigg[ -\frac{\Delta t}{4} \sum_{i,j} (L^{-1})^{ij}(\bm{X}_{\mathrm{m}})
 \notag
 \\[3pt]
 & \qquad \times \bigg[ \frac{X^{i}_{\mathrm{f}}-X^{i}_{\mathrm{i}}}{\Delta t}-\widetilde{\mathcal{J}}^{i}(\bm{X}_{\mathrm{m}})\bigg] \bigg[ \frac{X^{j}_{\mathrm{f}}-X^{j}_{\mathrm{i}}}{\Delta t} -\widetilde{\mathcal{J}}^{j}(\bm{X}_{\mathrm{m}})\bigg]
 \notag
 \\[3pt]
 & \qquad -\frac{\Delta t}{2}\sum_{i}\frac{\partial \mathcal{J}^{i}_{\mathrm{rev}}(\bm{X}_{\mathrm{m}})}{\partial X^{i}_{\mathrm{m}}} -\frac{\Delta t}{2}\sum_{i}\frac{\partial \mathcal{J}^{i}_{\mathrm{irr}}(\bm{X}_{\mathrm{m}})}{\partial X^{i}_{\mathrm{m}}} 
 \notag
 \\[3pt]
 & \qquad +\frac{\Delta t}{4}\sum_{i,j} \frac{\partial^{2}L^{ij} (\bm{X}_{\mathrm{m}})}{\partial X^{i}_{\mathrm{m}} \partial X^{j}_{\mathrm{m}}}\Bigg]
 \label{eq:PI_str_m}
\end{align}
with
\begin{align}
 \widetilde{\mathcal{J}}^{i} (\bm{X}_{\mathrm{m}}) &\equiv \mathcal{J}^{i}_{\mathrm{rev}} (\bm{X}_{\mathrm{m}}) + \mathcal{J}^{i}_{\mathrm{irr}} (\bm{X}_{\mathrm{m}}) - \sum_{j} \frac{\partial L^{ij} (\bm{X}_{\mathrm{m}})}{\partial X^{j}_{\mathrm{m}}} .
 \label{eq:def_Jtilde}
\end{align}
We also obtain
\begin{align}
 \mathcal{P}_{\Delta t}(\bm{X}_{\mathrm{i}}^{*}\vert\bm{X}_{\mathrm{f}}^{*}) & = \frac{1}{\sqrt{(4\pi \Delta t)^{N} \det \mathsf{L}_{\mathrm{T}} (\bm{X}_{\mathrm{m}})}} \exp \Bigg[ -\frac{\Delta t}{4} \sum_{i,j} (L^{-1}_{\mathrm{T}})^{ij}(\bm{X}_{\mathrm{m}})
 \notag
 \\[3pt]
 & \qquad \times \bigg[ \frac{X^{i}_{\mathrm{f}}-X^{i}_{\mathrm{i}}}{\Delta t}-\widetilde{\mathcal{J}}^{i}_{\mathrm{T}}(\bm{X}_{\mathrm{m}})\bigg] \bigg[ \frac{X^{j}_{\mathrm{f}}-X^{j}_{\mathrm{i}}}{\Delta t} -\widetilde{\mathcal{J}}^{j}_{\mathrm{T}}(\bm{X}_{\mathrm{m}})\bigg]
 \notag
 \\[3pt]
 & \qquad +\frac{\Delta t}{2}\sum_{i}\frac{\partial \mathcal{J}^{i}_{\mathrm{rev}}(\bm{X}_{\mathrm{m}})}{\partial X^{i}_{\mathrm{m}}} -\frac{\Delta t}{2}\sum_{i}\frac{\partial \mathcal{J}^{i}_{\mathrm{irr}}(\bm{X}_{\mathrm{m}})}{\partial X^{i}_{\mathrm{m}}} 
 \notag
 \\[3pt]
 & \qquad +\frac{\Delta t}{4}\sum_{i,j} \frac{\partial^{2}L^{ij}_{\mathrm{T}} (\bm{X}_{\mathrm{m}})}{\partial X^{i}_{\mathrm{m}} \partial X^{j}_{\mathrm{m}}}\Bigg]
 \label{eq:PI_str_m_back}
\end{align}
with
\begin{align}
 \widetilde{\mathcal{J}}^{i}_{\mathrm{T}} (\bm{X}_{\mathrm{m}}) &\equiv \mathcal{J}^{i}_{\mathrm{rev}} (\bm{X}_{\mathrm{m}}) - \mathcal{J}^{i}_{\mathrm{irr}} (\bm{X}_{\mathrm{m}}) + \sum_{j} \frac{\partial L^{ij}_{\mathrm{T}} (\bm{X}_{\mathrm{m}})}{\partial X^{j}_{\mathrm{m}}} .
\end{align}
Substituting (\ref{eq:PI_str_m}) and (\ref{eq:PI_str_m_back}) into (\ref{eq:DB_m}) with (\ref{eq:dist_m}), we obtain
\begin{align}
 & \frac{\Delta t}{4} \sum_{i,j} (L^{-1})^{ij}(\bm{X}_{\mathrm{m}}) \bigg[ \frac{X^{i}_{\mathrm{f}}-X^{i}_{\mathrm{i}}}{\Delta t}-\widetilde{\mathcal{J}}^{i}(\bm{X}_{\mathrm{m}})\bigg] \bigg[ \frac{X^{j}_{\mathrm{f}}-X^{j}_{\mathrm{i}}}{\Delta t} -\widetilde{\mathcal{J}}^{j}(\bm{X}_{\mathrm{m}})\bigg]
 \notag
 \\[3pt]
 & -\frac{\Delta t}{4} \sum_{i,j} (L^{-1}_{\mathrm{T}})^{ij}(\bm{X}_{\mathrm{m}}) \bigg[ \frac{X^{i}_{\mathrm{f}}-X^{i}_{\mathrm{i}}}{\Delta t}-\widetilde{\mathcal{J}}^{i}_{\mathrm{T}}(\bm{X}_{\mathrm{m}})\bigg] \bigg[ \frac{X^{j}_{\mathrm{f}}-X^{j}_{\mathrm{i}}}{\Delta t} -\widetilde{\mathcal{J}}^{j}_{\mathrm{T}}(\bm{X}_{\mathrm{m}})\bigg]
 \notag
 \\[3pt]
 & + \Delta t \sum_{i}\frac{\partial \mathcal{J}^{i}_{\mathrm{rev}}(\bm{X}_{\mathrm{m}})}{\partial X^{i}_{\mathrm{m}}} -\frac{\Delta t}{4}\sum_{i,j} \frac{\partial^{2}}{\partial X^{i}_{\mathrm{m}} \partial X^{j}_{\mathrm{m}}} \left[ L^{ij} (\bm{X}_{\mathrm{m}}) - L^{ij}_{\mathrm{T}} (\bm{X}_{\mathrm{m}})\right]
 \notag
 \\[3pt]
 & + \frac{1}{2}\log \frac{\det \mathsf{L} (\bm{X}_{\mathrm{m}})}{\det \mathsf{L}_{\mathrm{T}} (\bm{X}_{\mathrm{m}})} + \Delta t \sum_{i} \frac{X^{i}_{\mathrm{f}}-X^{i}_{\mathrm{i}}}{\Delta t} \frac{\partial S(\bm{X}_{\mathrm{m}})}{\partial X^{i}_{\mathrm{m}}} = 0,
 \label{eq:pre_FDT_m}
\end{align}
where we have used
\begin{equation}
 S(\bm{X}_{\mathrm{f}}) - S(\bm{X}_{\mathrm{i}}) = \Delta t \sum_{i} \frac{X^{i}_{\mathrm{f}}-X^{i}_{\mathrm{i}}}{\Delta t} \frac{\partial S(\bm{X}_{\mathrm{m}})}{\partial X^{i}_{\mathrm{m}}}+ O\left( (\Delta t)^{2}\left( \frac{X^{i}_{\mathrm{f}}-X^{i}_{\mathrm{i}}}{\Delta t} \right)^{2}\right) ,
\end{equation}
and where the $O\left((X^{i}_{\mathrm{f}}-X^{i}_{\mathrm{i}} )^{2}\right)$ terms in (\ref{eq:pre_FDT_m}) are irrelevant in the limit $\Delta t/\tau_{\rm macro} \to 0$.
Note that $\partial S(\bm{X})/\partial X^{i}$ are called the thermodynamic forces.
Because (\ref{eq:pre_FDT_m}) holds for any $X^{i}_{\mathrm{f}}-X^{i}_{\mathrm{i}}$ and $X^{i}_{\mathrm{m}}$, comparing the quadratic terms in $X^{i}_{\mathrm{f}}-X^{i}_{\mathrm{i}}$ in (\ref{eq:pre_FDT_m}) yields
\begin{equation}
 L^{ij}(\bm{X}) = L^{ij}_{\mathrm{T}}(\bm{X}).
  \label{eq:re_L}
\end{equation}
Comparing the first-order terms in $X^{i}_{\mathrm{f}}-X^{i}_{\mathrm{i}}$ in (\ref{eq:pre_FDT_m}) with (\ref{eq:re_L}), we also have
\begin{align}
 \mathcal{J}^{i}_{\mathrm{irr}} (\bm{X}) &= \sum_{j} L^{ij}(\bm{X}) \frac{\partial S(\bm{X})}{\partial X^{j}}+ \sum_{j} \frac{\partial L^{ij} (\bm{X})}{\partial X^{j}},
 \label{eq:re_Jirr}
\end{align}
which is called the fluctuation dissipation relation of the second kind.
Comparing the zero-order terms in $X^{i}_{\mathrm{f}}-X^{i}_{\mathrm{i}}$ in (\ref{eq:pre_FDT_m}) with (\ref{eq:re_L}) and (\ref{eq:re_Jirr}), we finally obtain
\begin{equation}
 \sum_{i}\frac{\partial}{\partial X^{i}}\left[ \mathcal{J}^{i}_{\mathrm{rev}} (\bm{X}) P_{\mathrm{eq}}(\bm{X})\right] =0.
  \label{eq:re_Jrev}
\end{equation}

% equation 

Now, we go back to (\ref{eq:TP_ito_m}).
This is interpreted as the transition probability for the discrete time Langevin equation (\ref{eq:L_multi}).
By taking the limit $\Delta t/t_{\rm macro} \to 0$, we obtain
\begin{align}
 \frac{\mathrm{d}X^{i}_{t}}{\mathrm{d}t} = \mathcal{J}^{i}_{\mathrm{rev}} (\bm{X}_{t}) + \sum_{j} L^{ij}(\bm{X}_{t}) \frac{\partial S(\bm{X}_{t})}{\partial X^{j}_{t}} + \sum_{j} \frac{\partial L^{ij} (\bm{X}_{t})}{\partial X^{j}_{t}} + \sum_{j} l^{ij} (\bm{X}_{t}) \cdot \xi^{j}_{t} ,
 \label{eq:NLE_str_m}
\end{align}
where we have used (\ref{eq:re_Jirr}), and $l^{ij}$ satisfies 
\begin{equation}
 L^{ij}(\bm{X}_{t}) = \frac{1}{2}\sum_{k} l^{ik} (\bm{X}_{t}) l^{jk} (\bm{X}_{t}).
\end{equation}
It should be noted that the third term on the right-hand side of (\ref{eq:NLE_str_m}) is not eliminated even if we replace $l^{ij} (\bm{X}_{t}) \cdot \xi^{j}_{t}$ by $l^{ij} (\bm{X}_{t}) \odot \xi^{j}_{t}$ for general multi-component systems.
We emphasize that the result (\ref{eq:NLE_str}) in the previous section is rather accidental.

\section{Concluding Remarks}

% brief summary 

In this paper, we have derived a universal form of nonlinear, multiplicative Langevin equations for slow variables in equilibrium systems.
The result is essentially equivalent to that derived by Green in 1952.
In contrast to previous studies, we first assume the separation of time scales.
Then, by using the central limit theorem, we can formalize the asymptotic form of the probability density for the time-averaged fluxes, which determines the time evolution of the slow variables due to the time-reversal symmetry of fluctuation.

% LD

Here, we refer to the large deviation theory.
Our assumption (\ref{eq:LDP_F}) means large deviation property for the conditional probability density which is more general assumption than the quadratic form of the large deviation function (\ref{eq:CLT}). 
For instance, when one derives a stochastic time evolution equation with white Poisson noises from a microscopic mechanical system, the large deviation property should be valid.
The validity condition for (\ref{eq:CLT}) depends on details of a microscopic mechanical system.
In equilibrium systems, a large deviation function of thermodynamic variables has all information about fluctuations of the thermodynamic variables, and is expressed in terms of a corresponding thermodynamic function, which can be derived from a microscopic mechanical system by using equilibrium statistical mechanics.
A central limit theorem can be derived from the large deviation function, and only gives a corresponding fluctuation-dissipation theorem.
In this sense, the large deviation is a more fundamental concept for analyzing fluctuations although physical aspects of the the large deviation function for the time-averaged flux remain to be studied.

% physics

Before ending this paper, we summarize future problems related to our results.
First, although discussions of physical phenomena are out of scope of this paper, it seems interesting to find a system in which the third term on the right-hand side of (\ref{eq:NLE_str_m}) plays an important role in phenomena.
We do not know any such examples explicitly, but there are many cases where transportation coefficients strongly depend on thermodynamic variables.
However, it should be noted that the contribution of this term is higher-order in macroscopic systems, as discussed in Section~\ref{sec:intro}.
We thus seek such systems among small systems or singular systems.

% hydrodynamic fluctuation

Second, we believe that the result and its derivation method may provide the final answer for formally describing slow variables in equilibrium systems.
For example, in principle, fluctuating hydrodynamics in equilibrium systems can be studied explicitly using the same method.
In this case, we consider long-wavelength fluctuations of locally conserved quantities, which are called hydrodynamic modes, to be slow variables.
Although the formulation may be developed similarly to the ideas in this paper, there will be many technical difficulties in performing concrete calculations.
We thus conjecture that a formal derivation of the Navier--Stokes equation from Hamiltonian particle systems~\cite{Sasa2014} may be helpful for completing this problem.
Related to the argument of fluctuating hydrodynamics, we recall the assumption in Section~\ref{sec:NLE_BP} that the relaxation time of the momentum is much larger than the time scales of the other degrees of freedom.
There are cases where this assumption does not hold when the time scale of hydrodynamic modes is comparable with that of the momentum of a Brownian particle, as observed in recent experiments~\cite{HuangETAL,FranoschETAL,KheifetsETAL}.
Deriving the Langevin equation describing the motion observed in these experiments is also a challenging task.

% out of equilibrium

Third, the obvious problem we should study in future is formulating the stochastic evolution of slow variables in systems out of equilibrium.
If the time scales are separated so that the slow variables are clearly defined, then the concept of large deviation can be used even in nonequilibrium cases.
Of course, there are many nonequilibrium phenomena in which a complete set of slow variables is not identified.
Although these have interesting phenomena, we do not have a systematic method for studying them.
Putting such systems aside, we focus on systems where the slow variables are defined.
If the Gaussian form of the large deviation function is effective, then the dynamics of the slow variables will be described by a Langevin equation.
Even for this simple class, we do not have a general form, because the symmetry property (\ref{eq:DB_m}) cannot be used.
Rather, one may find that difficulties already appear in writing a deterministic equation for slow variables before considering stochastic systems.
Nevertheless, our method will be applied to the Brownian motion in nonuniform temperatures where the system satisfies the detailed balance condition as studied in~\cite{Polettini}.

% direction

Finally, recent work explicitly derived deterministic order parameter equations near the order-disorder transition of the globally coupled XY model and the synchronization-desynchronization transition of the Kuramoto model~\cite{Sasa2015}.
The characteristic feature of the derivation method here is to use nonequilibrium identities such as the Jarzynski equality~\cite{Jarzynski} and the Hatano--Sasa equality~\cite{Hatano-Sasa}.
Although we need to assume a rather special type of probability distribution at the initial time, the calculation steps are substantially reduced.
We thus expect that the unified framework of the approach in~\cite{Sasa2015} and the theory developed in this paper may be the first step in the universal description of the stochastic evolution of slow variables out of equilibrium. 

\begin{acknowledgements}
The authors would like to thank K.~Kanazawa, K.~Kawaguchi, Y. Nakayama, and M.~Ueda for useful discussions.
The present study was supported by KAKENHI Nos. 22340109 and 25103002.
\end{acknowledgements}

\appendix\normalsize
\renewcommand{\theequation}{\Alph{section}.\arabic{equation}}
\setcounter{equation}{0}
\makeatletter
  \def\@seccntformat#1{%
    \@nameuse{@seccnt@prefix@#1}%
    \@nameuse{the#1}%
    \@nameuse{@seccnt@postfix@#1}%
    \@nameuse{@seccnt@afterskip@#1}}
  \def\@seccnt@prefix@section{Appendix }
  \def\@seccnt@postfix@section{:}
  \def\@seccnt@afterskip@section{\ }
  \def\@seccnt@afterskip@subsection{\ }
\makeatother

\section{Transition Probability for a Discretized Langevin Equation} \label{sec:PIaL}

\subsection{Normalization Condition}\label{subsec:NC}

We verify the normalization condition
\begin{align}
 \int \mathrm{d}x_{n+1}\; \mathcal{P}(x_{n+1}\vert x_{n}) = 1+o(dt)
 \label{eq:NC}
\end{align}
for the transition probability
\begin{align}
 \mathcal{P}(x_{n+1}\vert x_{n}) &= \frac{1}{\sqrt{4\pi dt G(\bar{x}_{n})}} \exp \Bigg[ -\frac{dt}{4G(\bar{x}_{n})} \left[ \frac{dx_{n}}{dt}-f(\bar{x}_{n}) + 2\alpha G' (\bar{x}_{n})\right]^{2}
 \notag
 \\[3pt]
 & \qquad - \alpha f'(\bar{x}_{n}) dt + \alpha^{2} G''(\bar{x}_{n}) dt\Bigg] ,
 \label{eq:PI_ito_a}
\end{align}
where primes denote derivatives with respect to the argument, $\bar{x}_{n}\equiv \alpha x_{n+1}+(1-\alpha) x_{n}$, and $dx_{n}\equiv x_{n+1}-x_{n}$.
Changing the variable of integration from $x_{n+1}$ to $z\equiv dx_{n}/\sqrt{dt}$, we obtain
\begin{align}
 & \int \mathrm{d}x_{n+1}\; \mathcal{P}(x_{n+1}\vert x_{n})
 \notag
 \\
 & \qquad = \int \mathrm{d}z \; \frac{1}{\sqrt{4\pi G(\bar{x}_{n})}} \exp \Bigg[ -\frac{dt}{4G(\bar{x}_{n})} \left[ \frac{z}{\sqrt{dt}}-f(\bar{x}_{n}) + 2\bar{\alpha} G'(\bar{x}_{n}) \right]^{2}
 \notag
 \\
 & \qquad \qquad -\alpha f'(\bar{x}_{n}) dt + \alpha^{2} G''(\bar{x}_{n}) dt\Bigg]
 \label{eq:NC_CV}
\end{align}
with $\bar{x}_{n}=x_{n}+\alpha z\sqrt{dt}$.
In this subsection, for any function $A(\cdot)$, we abbreviate $A(x_{n})$ to $A$.
Expanding the integrand on the right-hand side of (\ref{eq:NC_CV}) in powers of $\sqrt{dt}$, we have
\begin{align}
 & \frac{\exp \left[ -\dfrac{dt}{4G(\bar{x}_{n})} \left[ \dfrac{z}{\sqrt{dt}}-f(\bar{x}_{n}) + 2\alpha  G'(\bar{x}_{n}) \right]^{2} -\alpha f'(\bar{x}_{n}) dt + \alpha^{2} G''(\bar{x}_{n}) dt\right]}{\sqrt{4\pi G(\bar{x}_{n})}}
 \notag
 \\
 & \quad = \frac{\mathrm{e}^{-z^{2}/(4G)}}{\sqrt{4\pi G}} \Bigg\{ 1+\sqrt{dt}\left[ \frac{f}{2G}z-\frac{3\alpha G'}{2G}z+\frac{\alpha G'}{4G^{2}}z^{3}\right] +\frac{dt}{2} \bigg[ 
 \notag
 \\
 & \quad \qquad -(f-2\alpha G')\frac{3\alpha G'}{2G^{2}} \left( z^{2}-\frac{z^{4}}{6G}\right) -\frac{(f-2\alpha G')^{2}}{2G}\left( 1-\frac{z^{2}}{2G}\right)
 \notag
 \\
 & \quad \qquad -2\alpha f' \left( 1-\frac{z^{2}}{2G}\right) + \frac{3(\alpha G')^{2}}{4G^{2}}\left( z^{2}-\frac{z^{4}}{G}+\frac{z^{6}}{12G^{2}}\right)
 \notag
 \\
 & \quad \qquad +2\alpha^{2} G'' \left( 1-\frac{5z^{2}}{4G}+\frac{z^{4}}{8G^{2}}\right) \bigg] + o(dt) \Bigg\} .
 \label{eq:PI_exp}
\end{align}
Substituting (\ref{eq:PI_exp}) into (\ref{eq:NC_CV}) and using
\begin{align}
 & \int \mathrm{d}z\; \frac{\mathrm{e}^{-z^{2}/(4G)}}{\sqrt{4\pi G}}z^{2n} = (2n-1)!! (2G)^{n},
 \label{eq:Gauss_even}
 \\[2pt]
 & \int \mathrm{d}z\; \frac{\mathrm{e}^{-z^{2}/(4G)}}{\sqrt{4\pi G}}z^{2n-1} = 0,
 \label{eq:Gauss_odd}
\end{align}
for $n\in \mathbb{N}$, we obtain the normalization condition~(\ref{eq:NC}).

\subsection{Derivation of the Fokker--Planck Equation from the Transition Probability} \label{subsec:PI_FP}

To establish the equivalence of the Fokker--Planck equation and the path integral formulation, we derive the Fokker--Planck equation
\begin{equation}
 \frac{\partial}{\partial t}P(\chi ,t) = -\frac{\partial}{\partial \chi} \bigg[ f(\chi) P(\chi,t)\bigg] + \frac{\partial^{2}}{\partial \chi^{2}} \bigg[ G(\chi)P(\chi,t)\bigg]
  \label{eq:FP_ito}
\end{equation}
from the transition probability~(\ref{eq:PI_ito_a}) in the limit $dt\to 0$.
To arrive at (\ref{eq:FP_ito}), we evaluate
\begin{equation}
 P(x,n+1) = \int \mathrm{d}y\; \mathcal{P}(x\vert y) P(y,n).
\end{equation}
Changing the variable of integration from $y$ to $z=(x-y)/\sqrt{dt}$, we obtain
\begin{align}
 P(x,n+1) &= \int \mathrm{d}z \frac{P(x-z\sqrt{dt},n)}{\sqrt{4\pi G(\bar{x})}} \exp \Bigg[ -\frac{dt}{4G(\bar{x})} \left[ \frac{z}{\sqrt{dt}}-f(\bar{x}) + 2\alpha G'(\bar{x}) \right]^{2}
 \notag
 \\[3pt]
 & \qquad -\alpha f'(\bar{x}) dt + \alpha^{2} G''(\bar{x}) dt\Bigg]
  \label{eq:FP_app_bef_expa}
\end{align}
with $\bar{x}=x+(\alpha-1)z\sqrt{dt}$.
In this subsection, we abbreviate $A(x)$ and $A(x,n)$ to $A$ for any function $A$.
Expanding the right-hand side of (\ref{eq:FP_app_bef_expa}) in powers of $\sqrt{dt}$, we have
\begin{align}
 P(x,n+1) &= \int \mathrm{d}z\; \frac{\mathrm{e}^{-z^{2}/(4G)}}{\sqrt{4\pi G}} \left[ P-P'z\sqrt{dt}+P''\frac{z^{2}dt}{2}+o(dt)\right]
 \notag
 \\[3pt]
 &\qquad \times \Bigg\{ 1+\sqrt{dt}\left[ \frac{f-2\alpha G'}{2G}z-\frac{(\alpha -1)G'}{2G}z+\frac{(\alpha -1)G'}{4G^{2}}z^{3}\right]
 \notag
 \\[3pt]
 & \qquad +\frac{dt}{2} \bigg[ -(f-2\alpha G')\frac{3(\alpha -1)G'}{2G^{2}} \left( z^{2}-\frac{z^{4}}{6G}\right)
 \notag
 \\[3pt]
 & \qquad -\frac{(f-2\alpha G')^{2}}{2G}\left( 1-\frac{z^{2}}{2G}\right) -2\alpha f' \left( 1-\frac{(\alpha -1)z^{2}}{2\alpha G}\right)
 \notag
 \\[3pt]
 & \qquad  + \frac{3[(\alpha -1)G']^{2}}{4G^{2}}\left( z^{2}-\frac{z^{4}}{G}+\frac{z^{6}}{12G^{2}}\right)
 \notag
 \\[3pt]
 & \qquad +2\alpha ^{2} G'' \left( 1-\frac{(5\alpha -1)(\alpha -1)z^{2}}{4\alpha ^{2}G}+\frac{(\alpha -1)^{2}z^{4}}{8\alpha ^{2}G^{2}}\right) \bigg] + o(dt) \Bigg\}
 \notag
 \\[3pt]
 &=\left( 1-f'dt+G''dt\right) P - \left( fdt -2G'dt\right) P' + G dt P'' + o(dt).
\end{align}
Thus, we obtain
\begin{equation}
 \frac{P(x,n+1)-P(x,n)}{dt}=-\frac{\partial}{\partial x}\bigg[ f(x)P(x,n)\bigg] + \frac{\partial}{\partial x^2} \bigg[ G(x)P(x,n) \bigg] +\frac{o(dt)}{dt}.
\end{equation}
In the limit $dt \to 0$, we arrive at the Fokker--Planck equation~(\ref{eq:FP_ito}).

%%%%%%%%%%%%%%%%%%%%%%%
% References          %
%%%%%%%%%%%%%%%%%%%%%%%

\end{document}